\documentstyle[12pt]{article}

\newcommand{\bls}[1]{\renewcommand{\baselinestretch}{#1}}

\input{psfig}

\newcommand{\Logo}[1]{
\hspace{7.2cm}
\parbox{0cm}
{\psfig{figure=sugplogo.pp,height=2cm,bbllx=0cm,bblly=0cm,bburx=4cm,bbury=4cm}}

\vspace{-0mm}\bls{1}
\begin{flushright}
{\small #1} \\ {\small\today}
\end{flushright}

\vspace{-10mm}\begin{center}\rule[0cm]{15.7cm}{.2mm}\end{center}
}

\newcommand{\Title}[1]{\vspace{0mm}
    \begin{center}{\bls{1.3}\Large\bf #1 \\ \bls{1}}\end{center} \vspace{5mm}
}
\newcommand{\Author}[2]{\bls{1.4}
    \begin{center}{{\large {#1}} \\} \bls{1.2} {\small\it #2 \\} \end{center}
}
\vspace{5cm}
\newcommand{\Abstract}[1]
{\bls{1.1}
\vspace{3mm}\begin{center}\large\bf Abstract.\end{center}
\par \small #1 \vspace*{\fill}
}

\def\C{{\mathchoice
{\setbox0=\hbox{$\displaystyle\rm C$}\hbox{\hbox to0pt
{\kern0.4\wd0\vrule height0.9\ht0\hss}\box0}}
{\setbox0=\hbox{$\textstyle\rm C$}\hbox{\hbox to0pt
{\kern0.4\wd0\vrule height0.9\ht0\hss}\box0}}
{\setbox0=\hbox{$\scriptstyle\rm C$}\hbox{\hbox to0pt
{\kern0.4\wd0\vrule height0.9\ht0\hss}\box0}}
{\setbox0=\hbox{$\scriptscriptstyle\rm C$}\hbox{\hbox to0pt
{\kern0.4\wd0\vrule height0.9\ht0\hss}\box0}}}}

\begin{document}
\begin{titlepage}

\Logo{SU-GP-93/6-2, DF/IST 6.93}

\Title{
COMPLEXIFICATION OF GAUGE THEORIES
}

\Author{R. Loll\footnote{address after Aug.93: Center for Gravitational Physics
and Geometry, Penn State University, University Park, PA 16802, USA}}
{Physics Department, Syracuse University, Syracuse, NY 13244, USA}

\Author{J.M. Mour\~ao\footnote{address Oct.93-Jul.94:
  Center for Gravitational Physics
and Geometry, Penn State University, University Park, PA 16802, USA}}
{Dept. F\'{\i}sica, Inst. Superior T\'{e}cnico, 1096 Lisboa, PORTUGAL}

\Author{J.N. Tavares}
{Dept. Matem\'atica Pura, Fac. Ci\^encias, Univ. Porto, 4000
Porto, PORTUGAL\\and\\
Dept. F\'{\i}sica, Inst. Superior T\'{e}cnico, 1096 Lisboa, PORTUGAL}

\Abstract{
For the case of a first-class constrained system with equivariant momentum map,
we study the conditions under which the double process of reducing
to the constraint surface and dividing out by the group of gauge
transformations $G$
is equivalent to the single process of dividing out the initial
phase space by the complexification $G_\C$
of $G$. For the particular case of a phase space action that is the lift of
a configuration space action, conditions are found
under which, in finite dimensions, the physical phase space of a
gauge system with first-class constraints is diffeomorphic to
a manifold imbedded in the physical configuration space of the complexified
gauge system. Similar conditions are shown to hold
in the infinite-dimensional example of Yang-Mills theories.
As a physical application we discuss the adequateness of using
holomorphic Wilson loop variables as (generalized) global coordinates
on the physical phase space of Yang-Mills theory.
}

%\Comment{PACS-0123456789}

%\end{document}
%
%       End of cover page
%
\end{titlepage}

\def\R{{\rm I\!R}}
\def\C{{\mathchoice
{\setbox0=\hbox{$\displaystyle\rm C$}\hbox{\hbox to0pt
{\kern0.4\wd0\vrule height0.9\ht0\hss}\box0}}
{\setbox0=\hbox{$\textstyle\rm C$}\hbox{\hbox to0pt
{\kern0.4\wd0\vrule height0.9\ht0\hss}\box0}}
{\setbox0=\hbox{$\scriptstyle\rm C$}\hbox{\hbox to0pt
{\kern0.4\wd0\vrule height0.9\ht0\hss}\box0}}
{\setbox0=\hbox{$\scriptscriptstyle\rm C$}\hbox{\hbox to0pt
{\kern0.4\wd0\vrule height0.9\ht0\hss}\box0}}}}
\font\fivesans=cmss10 at 4.61pt
\font\sevensans=cmss10 at 6.81pt
\font\tensans=cmss10
\newfam\sansfam
\textfont\sansfam=\tensans\scriptfont\sansfam=\sevensans\scriptscriptfont
\sansfam=\fivesans
\def\sans{\fam\sansfam\tensans}
\def\Z{{\mathchoice
{\hbox{$\sans\textstyle Z\kern-0.4em Z$}}
{\hbox{$\sans\textstyle Z\kern-0.4em Z$}}
{\hbox{$\sans\scriptstyle Z\kern-0.3em Z$}}
{\hbox{$\sans\scriptscriptstyle Z\kern-0.2em Z$}}}}

%\documentstyle{article}
%\begin {document}

\section {Introduction}

\newtheorem{theo}{Theorem}
\newtheorem{prop}{Proposition}
\newtheorem{defi}{Definition}
\newtheorem{cond}{Conditions}

Complex gauge theories, i.e. theories with a complex group of
gauge transformations, have recently appeared in several physical
applications. The hamiltonian (constraint) equations of $3+1$
general relativity simplify significantly when written in terms of the
$SL(2,\C)$-Ashtekar connection and its canonically conjugate variable, the
densitized triad \cite{Ash90}. On the other hand, it has been shown that
general relativity with a positive cosmological constant in $2+1$
dimensions corresponds to a Chern-Simons theory with gauge group $SL(2,\C)$
\cite{Witt88}.

Motivated by these examples,
we study the geometric properties of a wide class of complex gauge
theories obtained by the ``complexification" of real gauge theories.
Most of our rigorous analysis takes place in finite dimensions, where
the analogous setting is that of hamiltonian gauge models with
first-class constraints.
In the infinite-dimensional case we prove our main result for Yang-Mills
theories with a compact structure group.

Our investigation is purely kinematical and concerns properties
of the big and reduced phase and configuration spaces.
In the case of phase spaces that are cotangent bundles and where the action of
the group of gauge transformations is the lift of an action on configuration
space, it
is shown that under reasonable conditions the physical phase space of the
real, kinematic gauge theory $(Q, L_G, Q_{ph} = Q / L_G)$ is diffeomorphic
(see {\it Th.3} and {\it Th.5})
to an open submanifold ${\cal C}^{sat}/L_{G_\C}$ of the physical
configuration space $Q^\C_{ph}$ of the complex gauge theory $(Q^\C, L_{G_\C},
Q^\C_{ph})$, which is the complexification of $(Q, L_G, Q_{ph})$
(see {\it Def.1}).
This result is shown to hold also in the more general case of phase spaces
where the action of the group of gauge transformations $G$ possesses an
equivariant momentum map.

The action of the real group $G$ on the phase space $\cal P$
is extended to an action of $G_\C$ by
having the imaginary generators act appropriately in the directions
orthogonal to the constraint surfaces. Note that the action of $G_\C$ will
in general be symplectic only if we restrict it to the real subgroup $G$.

When the saturation ${\cal C}^{sat}$ (containing all points of $\cal P$
that can be reached from the constraint surface
$\cal C$ by a complex gauge transformation) is
dense in $Q^\C$ - which was the case in all examples studied -
the above-mentioned diffeomorphism implies that in order
to find the physical, reduced phase space for the system under
consideration, the double process of restricting to the constraint
surface ${\cal C} \subset {\cal P}$ and dividing out by the real gauge
transformations is equivalent to dividing out by the {\it complex}
gauge transformations. This is in accordance with the
expectation that the complex gauge orbits have twice the dimension of
the real gauge orbits.

An analogous result for Chern-Simons theory was proven in \cite{Hit}, namely,
that the cotangent bundle of the physical phase space of a Chern-Simons
theory with compact gauge group $K$ is diffeomorphic to a dense submanifold
in the physical phase space of the theory with complex gauge group $K_\C$
\cite{Witt88}.

The equivalence between the physical phase space ${\cal P}_{ph}$ of the
real gauge theory and the physical configuration space $Q^\C_{ph}$ of the
complex gauge theory may play an important role for both theories.
On the one hand, the quantization of the complex theory,
by analogy with the Palatini theories \cite{AshRom},
is expected to be facilitated by the existence of additional structures
in $Q^\C_{ph}$, induced from ${\cal P}_{ph}$.
It also allows us in principle to relate
the Hilbert spaces associated with the quantization of the complex
theory with the better-understood Hilbert spaces of
the quantization of the real theory (an example is given by the generalization
to a complex gauge group of the Chern-Simons theory in $2+1$ dimensions by
Witten \cite{Witt91}).
On the other hand, global (generalized) coordinates in
$Q^\C_{ph}$ may be used in ${\cal P}_{ph}$. This motivates the
use of ``holomorphic Wilson loop variables" as global
coordinates on the physical phase space of Yang-Mills theories
with real gauge groups, and gives further justification to the use of such
variables in general relativity, written in the Ashtekar variables (cf. Sec.5).

We do not address here the general question of how a dynamical principle
can be incorporated into our framework. Standard hamiltonians for gauge
theories may not be physically meaningful in the complexified theory.
For example, it is well-known that the usual Yang-Mills action
proportional to $Tr\,FF$ for a complex gauge group $G$ leads to a
non-positive energy. Note also that the theories we cited at the
beginning of the introduction are examples of so-called generally covariant
theories, whose particular properties render them meaningful in spite of
the presence of complex structures. Furthermore, as illustrated by the
example of $3+1$ gravity in terms of Ashtekar variables, it may be
necessary to complement the theory by a set of reality conditions,
projecting out the sector of physical states.

The present work is organized as follows. In Sec.2 we describe the class
of gauge theories under study. The main result relating
${\cal P}_{ph}$ and $Q^\C_{ph}$ (i.e. ${\cal P}/L_{G_\C}$)
is formulated and proven in Sec.3.
In Sec.4 we study three illustrative finite-dimensional examples, with
the groups of gauge transformations $G=\R^n$, $G=SO(n)$, and $G=U(n)$
respectively. The latter is an example of an action that is not the lift of
a configuration space action. Its physical phase space is a complex
Grassmann manifold.
In Sec.5 we demonstrate that our techniques are applicable on the
infinite-dimensional phase space of Yang-Mills theory on an arbitrary
three-dimensional compact and oriented manifold. We also comment on
how general relativity in terms of the Ashtekar variables can be
viewed in the same framework. In Sec.6 we present our conclusions.
\vskip1.5cm

\section {Gauge theories in the hamiltonian formalism}

In order to set the stage for the field theoretic application, we
first study the geometry of the analogous finite-dimensional
hamiltonian systems with symmetry which, in Dirac's terminology, are
gauge models with a set of first-class constraints.

Let the finite-dimensional manifold $Q$ be the ``big" configuration
space of such a gauge system, and let $L_G$ denote
a proper, but not necessarily free action
of a Lie group $G$ as the group of gauge transformations on $Q$.
(Recall that an action $L_G$ is called proper if the inverse images of compact
sets under the map $(g,x)\rightarrow (L_g x,x)$ are again compact.)
Assuming that $Q/L_G$ (possibly after
excluding singular orbits from $Q$) has the structure of a
differentiable manifold, this quotient space is known as the
{\it physical configuration space} of the system,
\begin{equation} Q_{ph} = Q / L_G.                     \label{qphy}
\end{equation}
The resulting triplet
\begin{equation}
(Q, L_G, Q_{ph})                    \label{gthe}
\end{equation}
will be called a {\it kinematic gauge theory}. In our current
investigation we will not address the question of how a physical,
gauge-invariant
dynamics can be introduced into this setting in a meaningful way.

The big phase space of the gauge theory (\ref{gthe}) is the cotangent
bundle ${\cal P} = T^*Q$, with the canonical symplectic form $\Omega$.
The gauge transformations of the big configuration space, $L_g$, $ g
\in G$, lift uniquely to symplectic ($\Omega$-preserving) gauge
transformations $\widetilde L_g$, $ g \in G$, of $\cal P$.
Note that our results also apply when the phase space $\cal P$
is not of the form of a cotangent bundle and/or the symplectic
action of the symmetry group $G$ is not the lift of an action on the
configuration space (c.f. Sec.4.3).

We will now introduce the notion of a {\it momentum map}
\cite{AbrMar,GuiSte,Arm2},
which is a useful tool in the abstract formulation of hamiltonian systems
with  symmetry. The components of the momentum map are just the conserved
quantities associated with that symmetry (for example, the components
of the angular momentum of a particle in the presence of rotational
symmetry). For gauge systems, the conserved quantities are the
first-class constraints, which are required to vanish for physical
configurations. The constraints define a submanifold in phase space,
the so-called constraint surface $\cal C$, which can alternatively be
described as the zero level set of the corresponding momentum map.

For the class of gauge systems with action $\widetilde L_G$ we
are considering, the momentum map $\mu:{\cal P}
\rightarrow {\cal G}^{*}$ (${\cal G}^*$ denoting the dual of the
Lie algebra $\cal G$ of $G$) always exists and is constructed as
follows: for each algebra element $\xi \in {\cal G}$, let
the vector field $\xi^{\cal P} \in {\cal X}\big({\cal P} \big)$ be the
infinitesimal generator of the action $\widetilde L_G$ associated to
$\xi$,

\begin {equation}
\xi^{\cal P}(p) = \frac {d}{dt} \mid_{t=0}
\widetilde L_{exp (t\xi)} \ p, \ \ \ \ \ \forall p
\in {\cal P} \ .        \label{fvfp}
\end {equation}
Each such $\xi^{\cal P}$ is a globally hamiltonian vector
field on $\cal P$, with hamiltonian function $\mu^{\xi}:{\cal P}
\rightarrow R$ given by
\begin {equation}
{\mu}^{\xi}(\alpha_q) = \alpha_q \big({\xi}^{Q}(q) \big), \quad \qquad
\forall \alpha_q =p \in T^{*}Q,
\qquad  \label{momm}
\end {equation}
where ${\xi}^{Q}$ is the infinitesimal
generator of gauge transformations on $Q$, and we have identified a
phase space point $p$ with the corresponding one-form $\alpha$ on $Q$,
where $\alpha_q\in T^*_q Q$ with $q=\tilde\pi (p)$ ($\tilde\pi$ denoting the
projection to the base space $Q$).
Here $\mu^\xi$ is of course the first-class constraint associated
with the generator $\xi$ of the group of gauge transformations.

We now collect the maps $\mu^{\xi}$ into a unique momentum
map $\mu:{\cal P} \rightarrow {\cal G}^{*}$ by defining, for all $p
\in {\cal P}$ and $\xi \in {\cal G}$

\begin {equation}
\mu(p) \bullet \xi := \mu^{\xi}(p),    \label{momm2}
\end {equation}
where $\bullet$ denotes the duality between
$\cal G$ and ${\cal G}^{*}$.
The momentum map (\ref{momm}), (\ref{momm2}) is $Ad^{*}$-equivariant
\cite{AbrMar}, i.e.

\begin {equation}
Ad^{*}_{g^{-1}} \circ {\mu} = {\mu} \circ \widetilde L_g \ \ \ \ \ \forall g
\in G.
\end {equation}
This statement implies that the first-class
constraints form a true Lie algebra with respect to the Poisson
brackets on $\cal P$. In more general cases of phase spaces and symmetry group
actions we will assume that the $G$-action has an $Ad^*$-equivariant momentum
map.

To obtain the physical phase space of the kinematic gauge theory
$(Q, L_G, Q_{ph})$, we must use the momentum map twice. We
first have to restrict the phase space $\cal P$ to
the constraint surface ${\cal C}$,
defined as the zero level set of the momentum map, \begin{equation}
{\cal C} = \{ p \in {\cal P} \ : \mu (p) = 0 \}. \quad  \label{const}
\end{equation}
Furthermore, by equivariance the components $\mu^{\xi}$ of the
momentum map generate
gauge transformations $\widetilde L_g$ on $\cal C$, hence to obtain
the physical phase space ${\cal P}_{ph}$ we have to perform the
quotient
\begin{equation}
{\cal P}_{ph} =  {\cal C} / \widetilde L_G. \label{phph}
\end{equation}

In the present paper we show that in some physically relevant
cases the double process in phase space of constraining to $\cal C$ and
quotienting out by the real gauge group action of $G$ is equivalent to
the one-stage process of quotienting out by an appropriate
action of the {\it complex} group $G_\C$.

Recall that for the setting described above and for the case that
the group $G$ acts freely on $\cal P$, the following reduction theorem
holds (see, for example, \cite{MarWei} and references therein; the present
formulation is taken from \cite{AbrMar}, Theorem 4.3.1):

\begin{theo}
Let $({\cal P},\Omega)$ be a symplectic manifold on which the
Lie group $G$ acts symplectically and let $\mu :{\cal P}\rightarrow
{\cal G}^*$ be an $Ad^*$-equivariant momentum map for this action.
Assume $x\in{\cal G}^*$ is a regular value of $\mu$
(i.e. for every $p\in\mu^{-1}(x)$, $d\mu_p$ is surjective), and that the
isotropy group $G_x$ under the $Ad^*$-action on ${\cal G}^*$ acts
freely and properly on $\mu^{-1}(x)$. Then
${\cal P}_x=\mu^{-1}(x)/\widetilde L_{G_x}$ has a
unique symplectic form $\omega_x$ with the property
$\pi_x^{*}\omega_x = i_x^{*}\Omega$, where $\pi_x: \mu^{-1}(x)
\rightarrow {\cal P}_x$ is the canonical projection and $i_x:
\mu^{-1} (x)\rightarrow \cal P$ is the inclusion.
\end{theo}

There is an analogous result if
$x\in{\cal G}^*$ is only a weakly regular value of $\mu$
(i.e. $\mu^{-1}(x)$ is a submanifold with $T_p\mu^{-1}(x)=Ker\,d
\mu_p$), the group $G_x$ therefore does not act freely,
the $G_x$-orbits in $\mu^{-1}(x)$ are all of the same type,
and hence the dimension of the
isotropy group $G_p$ constant for all points $p\in\mu^{-1}(x)$).
Moreover, if the $G$-action on $Q$
is proper and free, one can prove that \cite{Got}

\begin {equation}
{\cal P}_{ph}={\cal C}/\widetilde L_G \cong T^{*}(Q/ L_G)
\end {equation}
i.e. the reduced physical phase space
${\cal P}_{ph}={\cal C}/G$, given by the above theorem, is
symplectomorphic
to the cotangent space $T^{*}Q_{ph}$ of the reduced physical
configuration space $Q_{ph} = Q/L_G$. Under appropriate regularity conditions,
an analogous result holds for the case that the action is non-free
\cite{Mont}, the standard action of $SO(n)$ on $\R^n$ being the prototypical
example (cf. Sec.4.2).
\vskip1.5cm

\section{ Complexification of gauge theories}

A complex Lie group $H$ is a Lie group which at the same time
is a complex manifold \cite{Hoc}. The two structures are
related by demanding that the map
\begin{eqnarray*}
H \times H & \longrightarrow & H \\
(h_1, h_2) &\mapsto & h_1 h_2^{-1}
\end{eqnarray*}
be holomorphic.

So far we have considered the general case of triplets
$(Q, L_G, Q_{ph})$, without specifying whether the group $G$ is
real or complex. We will call such a gauge theory complex if
the group $G$ of gauge transformations is a complex Lie group. It is
convenient to introduce the concept of the ``complexification" of
a kinematic gauge theory:

\begin{defi}
The kinematic gauge theory $(Q^\C, L_{G_\C}, Q^\C_{ph})$ is called a
complexification of the theory $(Q, L_{G}, Q_{ph})$ iff it satisfies
the following three conditions:
\begin{enumerate}
\item
The complex configuration space is diffeomorphic to the real phase space,
\begin{equation}
Q^\C \stackrel{diff}{=} {\cal P}(\equiv T^*Q).             \label{a4a}
\end{equation}

\item
The complex gauge group can be uniquely written in the form
\begin{equation}
G_\C = G e^{i {\cal G}}.  \label{a4b}
\end{equation}
In case the group $G$ is compact, $G_\C$ is
called the {\it universal complexification} of $G$, and $G$ is its
maximal compact subgroup \cite{Hoc}.

\item
Using the diffeomorphism (\ref{a4a}),
the restriction of the complex action of $G_\C$ to its real subgroup
$G$ coincides with the lift of the $L_G$-action,
\begin{equation}
L_{G_\C \downarrow G} = \widetilde L_G.             \label{a4c}
\end{equation}
\end{enumerate}
\end{defi}

\noindent
This definition is motivated by the fact that the configuration space of a
Yang-Mills theory with complexified structure group
$K_\C$ satisfies
(\ref{a4a}), (\ref{a4b}) and (\ref{a4c}), and hence can be regarded as the
complexification of a Yang-Mills theory with structure group $K$ (cf. Sec.5).
- In the present paper we study the conditions under which the equation
(\ref{a4a}) holds at the level of the {\it reduced} spaces ({\it if} a
complexification of  $(Q, L_{G}, Q_{ph})$ exists), i.e. when we have
\begin{equation}
Q^\C_{ph}  \stackrel{diff}{=} {\cal P}_{ph},     \label{a5}
\end{equation}
where ${\cal P}_{ph}$ is defined by (\ref{phph}).

Consider now the (not necessarily unique) complexification
$(Q^\C, L_{G_\C}, Q^\C_{ph})$ of the kinematic gauge theory
$(Q, L_{G}, Q_{ph})$. The action of $G_\C$ on $\cal P$ defines
a homomorphism $\tau$ of Lie algebras from the complexified Lie algebra
${\cal G}_\C$ into the vector fields on $\cal P$,
\begin{eqnarray}
\tau \ : \ {\cal G}_\C =  {\cal G} + i  {\cal G} &\longrightarrow&
{\cal X}\big({\cal P} \big) \label{a7a} \nonumber \\
\tau\big(\xi + i \eta \big)  &=& \xi^{\cal P} +
\big( i \eta \big)^{\cal P},    \qquad \qquad \xi, \eta \in {\cal G}.
\qquad  \label{a7b}
\end{eqnarray}
We will assume that $\cal P$ and $L_{G_\C}$ are such that
the following conditions hold:

\begin{cond}
\begin{itemize}

\item There exists a symplectic almost-complex structure ${\bf J}$
on $\cal P$ such that
\begin{equation}
\big(i \eta \big)^{\cal P} = {\bf J} \eta^{\cal P}, \quad
\qquad \forall \eta \in {\cal G}.    \qquad  \label{a8}
\end{equation}

\item The non-degenerate symmetric tensor $\gamma$ on $\cal P$ defined
by $\Omega$ and $\bf J$ through
\begin{equation}
\gamma (X, Y) := \Omega(X, {\bf J} Y)      \label{a9}
\end{equation}
is a Riemannian metric, i.e. $\cal P$ is a quasi-K\"ahler manifold.

\end{itemize}
\end{cond}

There are a number of conclusions that can
be drawn from the existence on ${\cal P}$ of an almost-complex structure
$\bf J$ and a Riemannian metric $\gamma$ that intertwine with the
symplectic structure according to (\ref{a9}):

\begin{itemize}
\item For all linear subspaces $ S \subset T_p{\cal P}$ we have
\begin {equation}
{\bf J}(S^{\perp}) = S^{\circ}      \label{posym}
\end {equation}
where $S^{\perp}$ denotes the subspace ($\gamma$-)orthogonal and
$S^{\circ}$ the subspace polar symplectic (or, $\Omega$-orthogonal,)
to $S$.

\item Denoting by $\nabla {\mu}^{\xi}$ the gradient vector field with
respect to $\gamma$ of the constraint
function ${\mu}^{\xi}:{\cal P} \rightarrow R$,
\begin {equation}
d{\mu}^{\xi}(Y)= \gamma(\nabla {\mu}^{\xi},Y), \ \ \ \forall Y \in
{\cal X}\big({\cal P} \big),
\label{grad}
\end {equation}
we have
\begin {equation}
\nabla {\mu}^{\xi} = {\bf J}{\xi}^{\cal P},
\end {equation}
i.e. the vector space perpendicular to a surface $\mu=$const.
is obtained by applying the complex structure $\bf J$ to the
infinitesimal gauge generators $\xi^{\cal P}$.
In fact, using (\ref{grad}) and the fundamental relation (\ref{a9}),
one derives for all $Y \in {\cal X}\big({\cal P} \big)$
\begin {equation}
\gamma(\nabla{\mu}^{\xi},Y) = d{\mu}^{\xi}(Y)=\Omega({\xi}^{\cal P},Y)=
\gamma({\bf J}{\xi}^{\cal P},Y). \ \  \nonumber
\end {equation}

\item For each point $p \in {\cal P}$, define a map $\nabla
{\mu}_p:{\cal G} \rightarrow T_p{\cal P}$ by
\begin {equation}
\nabla {\mu}_p(\xi) = \nabla {\mu}^{\xi}(p) = {\bf J}{\xi}^{\cal P}(p).
\end {equation}
Identifying the tangent space $T_p{\cal P}$ with the
cotangent space $T^{*}_p{\cal P}$ via the Riemannian metric $\gamma$,
we see that the map $\nabla {\mu}_p:{\cal G} \rightarrow T_p{\cal P}$
is the adjoint of the map $d{\mu}_p:T_p{\cal P} \rightarrow {\cal
G}^{*}$.

\item It is easy to verify that
\begin {equation}
T_p(\widetilde L_G \cdot p) = (Ker \, d{\mu}_p)^\circ =
{\bf J}(Im {\nabla}{\mu}_p)
\end {equation}
and
\begin {equation}
Ker ({\nabla}{\mu}_p) = {\cal G}_p     \label{gorbi}
\end {equation}
where $\widetilde L_G \cdot p$
denotes the $G$-orbit through the point $p \in {\cal P }$, and
${\cal G}_p$ the Lie algebra of the isotropy group $G_p$ at $p$.
\end{itemize}

Thus, if the pair $(L_{G_\C},{\cal P})$ satisfies {\it Cond.1}, the
imaginary generators of $G_\C$ are represented by vector fields
$\gamma$-orthogonal to the surfaces $\mu=const$. Note that these vector
fields ${\bf J}\xi^{\cal P}$ are in general neither hamiltonian nor
isometries.

Note furthermore that although the $G$-action on $\cal P$ is proper, the
$G_\C$-action need not be, which implies that the quotient $Q^\C/L_{G_\C}$ need
not be Hausdorff. This of course can only occur since $G_\C$ is a
{\it non-compact} group. Although one can make sense of the case when the
$G$-action is proper and the quotient space $Q/L_G$ is an orbifold (i.e. no
longer a smooth manifold) \cite{Emm}, we are not aware of a treatment or a
physical interpretation for the non-Hausdorff case. Our work may be viewed as a
prescription of how to deal with such gauge systems, by selecting a
sufficiently
well-behaved subspace of $Q^\C$.

Let us now recall the Moncrief decomposition for gauge systems
\cite{Mon1,AFM} which characterizes an orthogonal splitting of
the tangent spaces $T_p{\cal P}$, where $p$ lies in the constraint
surface $\cal C$. Taking into account that
\begin{equation}
T_p(\widetilde L_G \cdot p)\subset Ker \, d{\mu}_p
\end{equation}
for all points $p\in\cal C$, (\ref{posym})-(\ref{gorbi}) can be
summarized into the following theorem.

\begin{theo}
At all points $p \in {\cal C} \subset {\cal P}$, the tangent space
$T_p{\cal P}$ admits
the following (orthogonal) Moncrief decomposition:
\begin{eqnarray}
T_p{\cal P} = \underbrace {Ker \, d{\mu}_p \cap {\bf J}(Ker \, d{\mu}_p)
}_{(\bf 1)} \oplus \underbrace {Im \nabla {\mu}_p}_{(\bf 2)} \oplus
\underbrace { {\bf J}(Im \nabla {\mu}_p)}_{(\bf 3)} \nonumber \\
{ }  \label{moncr}
\end{eqnarray}
the first summand being symplectic, and the last two isotropic.

\end{theo}

At a given point $p \in {\cal C}$, the Moncrief
decomposition (\ref{moncr}) of $T_p{\cal P}$ has the following geometric
interpretation:

\begin{itemize}

\item $({\bf 1})= Ker \, d{\mu}_p \cap {\bf J}(Ker \, d{\mu}_p)$ can be
naturally identified with the tangent space $T_{\pi(p)}{\cal P}_{ph}$,
and represents the {\it true, physical degrees of freedom} of the
system.

\item  $({\bf 2})=Im \nabla {\mu}_p$ represents infinitesimal
deformations orthogonal to the constraint surface $\cal C$.

\item  $({\bf 3})={\bf J}\big( Im \nabla {\mu}_p \big)=
T_p(\widetilde L_G \cdot p)$ is
the tangent space to the gauge directions, that is, to the
$G$-orbit $\widetilde L_G \cdot p$ through $p \in {\cal C}$.

\end{itemize}
Consider now the subset ${\cal C}^{sat}$ of $\cal P$ given by
\begin{equation}
{\cal C}^{sat} := \{L_g \cdot p \ : \ p \in {\cal C} \ and \
g \in G_\C \}.         \qquad  \label{csat}
\end{equation}
We will call the set ${\cal C}^{sat}$ the {\it saturation of $\cal C$
(with respect to the action of $G_\C$)}. It consists of all points that
can be reached from the constraint surface by a complex gauge
transformation.
We prove below that ${\cal C}^{sat}$ is open in $\cal P$
and conjecture that it is actually dense in $\cal P$.
In Sec.4 we corroborate this conjecture by some examples.

The Moncrief decomposition (\ref{moncr}) implies the ``local" (in a
neighbourhood of $\cal C$)
equivalence between $ {\cal C}^{sat} / L_{G_\C}$ and ${\cal P}_{ph}
= {\cal C} / \widetilde {L}_G$. The following propositions prove their
global equivalence.

\begin{prop}
The saturation ${\cal C}^{sat}$ is open in the phase space $\cal P$.

\bigskip

{\underline {Proof.}}  {\small By virtue of the Moncrief decomposition,
$\{{\xi}^{\cal P}_p : \xi \in i{\cal G}\} = Im {\nabla {\mu}_p}$ is a
complementary subspace to $T_p{\cal C}$ in $T_p{\cal P}$.
Therefore ${\cal C}^{sat}$ contains an open neighbourhood ${\cal
U}$ of $\cal C$ in $\cal P$. Moreover, since ${\cal C}^{sat}$ is
given by the union ${\cal C}^{sat} = \bigcup _{g \in G_\C} \{L_g \cdot
{\cal U}\}$, it follows that the saturation ${\cal C}^{sat}$ is
open in $\cal P$. QED.}
\end{prop}
 Now, since ${\cal C} \subset {\cal C}^{sat}$
 and since each ``real" orbit $ \widetilde L_G
\cdot p$, $p \in {\cal C}$ is contained in a ``complex" orbit $L_{G_\C}
\cdot p$, we have a map
 \begin {equation}
{\cal P}_{ph} = {\cal C}/ \widetilde L_G
\rightarrow {\cal C}^{sat}/L_{G_\C}.  \ \
\label{difm}
\end {equation}

Since our aim is to prove the equivalence between the physical phase
space and the complex quotient on the right-hand side, we must show
that the map (\ref{difm}) is a bijection. Equivalently, we must prove
that each ``complex"
orbit $L_{G_\C} \cdot p$, $p \in {\cal C}$, contains {\it only
one} ``real" orbit $L_G \cdot p$, i.e. that the following is true:

\begin{prop}
\begin {equation}
(L_{G_\C} \cdot p) \cap {\cal C} = L_G \cdot p, \ \ \ \ \forall p \in
{\cal C}. \end {equation}

\bigskip

{\underline {Proof.}} {\small It suffices to prove that
\begin {equation}
(L_{G_\C} \cdot p) \cap {\cal C} \subset L_G \cdot p.
\end {equation}
Let $L_g \cdot p \in ( L_{G_\C} \cdot p )\cap {\cal C}$, with $g \in
G_\C$. We must show that $L_g \cdot p = L_h \cdot p$, for some real
group element $h
\in G$. However, since $\cal C$ is a $G$-invariant submanifold of
$\cal P$ and $G_\C = G.exp (i{\cal G})$ by assumption, we can without
loss of generality take $g$ to be of the form $g=exp(i{\xi})$, with
$\xi \in {\cal G}$.

Hence assume that
\begin {equation}
L_{exp(i{\xi})} \cdot p = L_{e^{i\xi}} \cdot p \in {\cal C} \ .
\label{eixi}
\end {equation}

Let us set $q_t:=L_{e^{it\xi}} \cdot p$, and consider the
function $f:[0,1] \rightarrow R$, defined by (see equation (\ref{momm2}))
\begin {equation}
f(t) = {\mu}^{\xi} (q_t).
\end {equation}

Then, $f(0)={\mu}^{\xi}(p)=0$, since $p \in {\cal
C}={\mu}^{-1}(0)$, and also $f(1)={\mu}^{\xi}(q_1)=0$, by assumption
(\ref{eixi}).
On the other hand, using (\ref{a7b}) and (\ref{a9}) we compute

\begin{eqnarray}
f'(t) &=& d{\mu}^{\xi}(q_t) \big( {\bf J}{\xi}^{\cal P}(q_t) \big)
\nonumber \\
     &=& \Omega_{q_t} \big( {\xi}^{\cal P}(q_t),{\bf J}{\xi}^{\cal P}(q_t)
\big) \nonumber \\
     &=& \gamma({\xi}^{\cal P}(q_t),{\xi}^{\cal P}(q_t))
\geq 0.
\end{eqnarray}

Therefore, the function $f$ must vanish on the entire interval
$[0,1]$, and $q_t \in {\cal C}$, for all $t\in [0,1]$.
Moreover we have, for all $t\in [0,1]$, $\xi^{\cal P} (q_t) = 0$ which
implies $q_t =p$ and, in particular,
\begin{equation}
L_{e^{i\xi}} \cdot p = p \in L_G \cdot p.
\end {equation}
QED.}

\end{prop}

In conclusion, the map (\ref{difm})
is a bijection. Moreover, using the Moncrief
decomposition and the inverse mapping theorem, one can easily prove
that this map is in fact a ($\bf J$-dependent) diffeomorphism.
We have therefore proved our main result:

\begin {theo}
Let $\widetilde L_G$ be a proper symplectic action of the
Lie group G on the symplectic manifold {\cal P}, with
an $Ad^*$-equivariant momentum map $\mu$, and let
$L_{G_\C}$ be a complexification of this action
satisfying (\ref{a4b}), (\ref{a4c}) and {\it Cond.1}.
If $0$ is a weakly regular value of $\mu$, the map
(\ref {difm}) is a diffeomorphism between ${\cal C} / \widetilde L_G$
and ${\cal C}^{sat} / L_{G_\C}$, where ${\cal C}^{sat}$
is given by (\ref{csat}).

\end{theo}

Note that there may exist complex orbits in $Q^\C$ which do not
contain any real orbit $\widetilde L_G\cdot p,\; p\in\cal C$, i.e.
not every complex orbit necessarily intersects the
constraint surface (see examples 4.2 and 4.3).
However, in the finite-dimensional examples we considered, ${\cal C}^{sat}$ was
always dense in $\cal P$. We conjecture
that this is the case for a wide class of
systems and in particular for the Yang-Mills theory discussed in Sec.5.

Given the diffeomorphism between ${\cal P}_{ph}$ and ${\cal
C}^{sat}/L_{G_\C}$, we can now pull back to ${\cal
C}^{sat}/L_{G_\C}$ the unique symplectic form on the physical phase
space obtained from the Marsden-Weinstein reduction.

Conversely, we are interested in the question under what conditions
a given kinematic gauge theory $(Q, L_G, Q_{ph})$ can be complexified.
One necessary condition is that the vector fields ${\bf J}\eta^{\cal
P}$ (c.f. (\ref{a8})) be complete, so that their
infinitesimal action can be exponentiated. We also need a
$G$-invariant almost complex structure $\bf J$ on the phase space $\cal
P$. This is not a very strong restriction, since for compact $G$ one can
always define a $G$-invariant almost-K\"ahler structure on $\cal P$.
The most important condition comes from demanding that the map
$\tau :{\cal G}_\C\rightarrow {\cal X} ({\cal P})$ constructed according to
\begin{equation}
\tau (\xi + i\eta):= \xi^{\cal P} + {\bf J}\eta^{\cal P},
\end{equation}
with $\xi,\eta \in {\cal G}$, define a homomorphism of Lie algebras
(cf. expressions (\ref{a7b}) and (\ref{a8})). It turns out that this is the
case only if the restriction of the
almost-complex structure $\bf J$ to
the complex $G_\C$-orbits in $\cal P$ is integrable.
- All these conditions are satisfied for the finite-dimensional gauge
systems discussed in the next section.
\vskip1.5cm

\section {Finite-dimensional examples}

\subsection{Group of gauge transformations $\R^n$}

We first consider an example with an abelian group of gauge transformations,
the kinematic gauge
model $(Q,L_G,Q_{ph})=(\R^m, L_{\R^n}, \R^{m-n})$, where $n<m$, with
$\R^n$ acting freely on $\R^m$. The coordinates on the configuration space
$Q$ are the $x_I$, $I=1,\dots ,m$, and the $n$ algebra generators $\xi^i$
of the gauge group $\R^n$ are taken to act on $Q$ by the vector fields
$\partial/\partial x_i$, $i=1,\dots ,n$.
The corresponding first-class constraints
on ${\cal P}=T^*\R^m$ are given by $p_i=0$, $i=1,\dots ,n$, with the
canonically lifted action on $\cal P$,
\begin{equation}
\xi^i\mapsto \frac{\partial}{\partial x_i},\qquad i=1,\dots ,n.
\label{com1}
\end{equation}
The $(2m-n)$-dimensional constraint surface $\cal C$ is given by
\begin{equation}
{\cal C}=\{ (\vec x,\vec p)\in\R^{2m}\, :\, p_i=0,\, i=1,\dots ,n \}.
\end{equation}
The physical phase space ${\cal P}_{ph}$ is the cotangent bundle
$T^*\R^{m-n}$, parametrized by the $m-n$ coordinate pairs $(x_a,p_a)$,
$a=n+1,\dots ,m$, with the induced canonical symplectic form.

We are now looking for suitable complexifications of
$(\R^m, L_{\R^n}, \R^{m-n})$. Consider the complex structure $\bf J$ on
$\cal P$ defined by
\begin{equation}
{\bf J}\frac{\partial}{\partial x_I}=\frac{\partial}{\partial p_I},\qquad
{\bf J}\frac{\partial}{\partial p_I}=-\frac{\partial}{\partial x_I},\qquad
I=1,\dots ,m.
\label{jact}
\end{equation}
We use this complex structure to extend the action of $\R^n$ to one of
$(\R^n)_\C=\C^n$ by representing the ``imaginary" generators as
\begin{equation}
i\xi^j\mapsto {\bf J}\frac{\partial}{\partial x_j}=
\frac{\partial}{\partial p_j},\qquad
j=1,\dots ,n.
\label{com2}
\end{equation}
If we now consider the kinematic gauge theory $(\C^m,L_{\C^n},\C^{m-n})$
with coordinates $z_I=x_I+i p_I$ on the complex configuration space
$\C^m$, and with the action of the gauge group $\C^n$ defined by
\begin{equation}
\tau(\xi^i +i\xi^j)=\frac{\partial}{\partial x_i} +i\frac{\partial}{\partial
p_j
   },\qquad
i,j=1,\dots ,n,
\end{equation}
by {\it Def.1} this is a complexification of the triplet
$(\R^m,L_{\R^n},\R^{m-n})$, under the diffeomorphism $z_I\rightarrow
(x_I,p_I)$ between $\C^m$ and $T^*\R^m$. Moreover, the symmetric tensor
$\gamma$ on $\cal P$ constructed according to (\ref{a9}) is a well-defined
Riemannian metric,
\begin{eqnarray}
\gamma (\frac{\partial}{\partial x_I},\frac{\partial}{\partial x_J})&=&
  \Omega (\frac{\partial}{\partial x_I},\frac{\partial}{\partial p_J}) =
\delta_{IJ}
\nonumber \\
\gamma (\frac{\partial}{\partial x_I},\frac{\partial}{\partial p_J})&=&
  \Omega (\frac{\partial}{\partial x_I},-\frac{\partial}{\partial x_J}) = 0
\nonumber \\
\gamma (\frac{\partial}{\partial p_I},\frac{\partial}{\partial p_J})&=&
  \Omega (\frac{\partial}{\partial p_I},-\frac{\partial}{\partial x_J}) =
\delta_{IJ},
\end{eqnarray}
so that {\it Cond.1} are satisfied.
The saturation ${\cal C}^{sat}$ of $\cal C$ in $\cal P$ under the action of
$\C^n$, (\ref{com1}) and (\ref{com2}), is all of $\cal P$, and any
$\C^n$-orbit is of the form
\begin{equation}
\{ (\vec x,\vec p)\in\R^{2m}\, :\, x_i,p_i\, {\em fixed, } \,
i=n+1,\dots ,m \},
\end{equation}
and therefore contains exactly one $\R^n$-orbit of the constraint surface
$\cal C$, demonstrating the desired equivalence of the quotients
\begin{equation}
{\cal C}^{sat}/\C^n \cong {\cal C}/\R^n .
\end{equation}

Let us now slightly modify the definition of the complex structure on
$\cal P$ by defining
\begin{equation}
{\bf J}'\frac{\partial}{\partial x_1}=\frac{\partial}{\partial p_m},\qquad
{\bf J}'\frac{\partial}{\partial x_m}=\frac{\partial}{\partial p_1},\qquad
{\bf J}'\frac{\partial}{\partial p_1}=-\frac{\partial}{\partial x_m},\qquad
{\bf J}'\frac{\partial}{\partial p_m}=-\frac{\partial}{\partial x_1},\qquad
\end{equation}
and all other relations unchanged from (\ref{jact}). Proceeding as above
leads to a complex action on $\C^m$ with
\begin{equation}
\tau(i\xi^1)=i\frac{\partial}{\partial p_m}.
\end{equation}
A $\C^n$-orbit on $\cal P$ is now of the form
\begin{equation}
\{ (\vec x,\vec p)\in\R^{2m}\, :\, p_1, x_m\, {\em fixed}; \,
x_a,p_a\, {\em fixed,} \, a=n+1,\dots ,m-1 \}.
\end{equation}
Such an orbit does not intersect $\cal C$ unless $p_1=0$, in which case it
contains a whole one-parameter family (labelled by $p_m$)
of $\R^n$-orbits in $\cal C$, which
in turn are of the form
\begin{equation}
\{ (\vec x,\vec p)\in\R^{2m}\, :\, p_i=0,\,i=1,\dots ,n;  \,
x_a,p_a\,\, {\em fixed,} \, a=n+1,\dots ,m \}.
\end{equation}
However, this does not contradict our conjecture since the metric
$\gamma'$ constructed according to (\ref{a9}) is not Riemannian; its signature
is $(-,-,+,\dots,+)$.

\subsection{Group of gauge transformations $SO(n)$}

Let us now consider a typical non-abelian kinematic gauge theory,
given by \cite{sha}
\begin{equation}
(Q, L_G, Q_{ph}) = (\R^n, L_{SO(n)}, \R^+)  ,   \label{4.2.1}
\end{equation}
where $\R^+ = \{ r \in \R,  0 < r < \infty \}$, $L_{SO(n)}$ denotes the
action of $SO(n)$ in the fundamental representation,
\begin{eqnarray}
x \in \R^n \mapsto L_A x &=& A x   \nonumber \\
A^t A &=& Id  , \; det\, A = 1   ,    \label{4.2.2}
\end{eqnarray}
with $Id$ denoting the $n \times n$-identity matrix. The action
$L_{SO(n)}$ is non-free for $n > 2$ and the physical configuration
space is the one-dimensional manifold $\R^+$ (after excluding the
singular orbit $\{0 \}$). The canonical lift of $L_{SO(n)}$ to
the phase space is
\begin{eqnarray}
\widetilde L_A \ : {\cal P} & \rightarrow& {\cal P} \nonumber \\
\widetilde L_A (x, p) &=& (Ax, Ap) , \label{4.2.3}
\end{eqnarray}
where $(x, p) \in {\cal P} = T^* \R^n = \R^n \times \R^n$.

The equivariant momentum map is, in accordance with (\ref{momm}),
\begin{eqnarray}
\mu (x, p) (T_{jk}) &=& p_i dx_i (T_{jk}^Q) =    \nonumber   \\
                    &=& x T_{jk} p^t   \qquad 1 \leq j < k \leq n,
\label{4.2.4}
\end{eqnarray}
where
\begin{eqnarray}
T_{jk} &\in& so(n) ,  \nonumber  \\
(T_{jk})_{j'k'} &=& \delta_{jj'} \delta_{kk'}
- \delta_{jk'} \delta_{kj'}  \ ,   \nonumber \\
T_{jk}^Q &=& x_j \frac {\partial}{\partial x_k} -
 x_k \frac{\partial}{\partial x_j}  .    \nonumber
\end{eqnarray}
The first-class constraints  associated with (\ref{4.2.4}) take the form
\begin{equation}
x T_{jk} p^t = 0, \qquad 1 \leq j < k \leq n,\nonumber
\end{equation}
or, equivalently,
\begin{equation}
x \wedge p = 0 ,     \label{4.2.5}
\end{equation}
where the wedge in (\ref{4.2.5}) denotes the
exterior product of the two vectors
$x , p \in \R^n$ and we have identified $so(n)$ with
$\wedge^2 \R^n$.
 By exterior algebra arguments, (\ref{4.2.5})
implies that the constraint surface $\cal C$ is given by
\begin{equation}
{\cal C} = \{ (x, p) \in {\cal P}: (x, p) = (\lambda f, \mu f), \
\lambda, \mu \in \R,  f \in \R^n , \parallel  f \parallel^2 =
\sum_{i = 1}^n f_i^2 = 1 \}  . \label{4.2.6}
\end{equation}
To obtain the physical phase space, we must divide $\cal C$ by the action
of $\widetilde L_{SO(n)}$. Since the action of $SO(n)$ is transitive
on the unit sphere,
\begin{equation}
S^{n-1} = \{ f \in \R^n , \parallel f \parallel = 1
\} \subset \R^n,     \nonumber
\end{equation}
we have in each $\widetilde L_{SO(n)}-$orbit a representative
\begin{equation}
(\lambda f_1 , \mu f_1 )  \in {\cal C}   , \label{4.2.7}
\end{equation}
where $f_1 = (1,0, \ldots , 0)$. There is a residual gauge
transformation
\begin{equation}
(\lambda f_1 , \mu f_1 ) \mapsto (- \lambda f_1 , - \mu f_1 ) ,
  \label{4.2.8}
\end{equation}
which means that the physical phase space is a cone \cite{sha}:
\begin{eqnarray}
{\cal P}_{ph} = {\cal C} / \widetilde L_{SO(n)} &=& \nonumber \\
= \R^2 / \Z_2 &=&
\{ [(\lambda ,\mu)] :  (\lambda , \mu) \in \R^2 \} ,
\label{4.2.9}
\end{eqnarray}
where we denote the $\Z_2$-equivalence classes by
$[( \lambda , \mu) ] = \{ (\lambda , \mu) , (- \lambda , - \mu) \}$.
Alternatively, if we exclude the origin of $\R^{2n}$ and choose
\begin{equation}
{\cal P} = \R^{2n}_* = \R^{2n} \backslash \{ 0 \} , \label{4.2.10}
\end{equation}
${\cal P}_{ph}$ becomes a non-simply connected two-dimensional manifold,
\begin{equation}
{\cal C} / \widetilde L_{SO(n)}  \equiv \R^2_{*} = \R^2 \backslash \{ 0 \} .
\label{4.2.11a}
\end{equation}

Let us now see how this process of constraining and quotienting
can be replaced by the single step of
quotienting the big phase space (\ref{4.2.10})
by the action of the complex group
\begin{equation}
SO(n)_\C = SO(n , \C) . \label{4.2.11b}
\end{equation}
Consider the complex kinematic gauge theory
\begin{equation}
(Q^\C , L_{G_\C} , Q^\C_{ph} ) = ( \C^n , L_{SO(n, \C)} ,
\C^n / L_{SO(n, \C)}) , \label{4.2.12}
\end{equation}
where $L_{SO(n, \C)}$ denotes
the standard (non-symplectic and, in fact, improper) action of $SO(n, \C)$ on
$\C^n$,
\begin{eqnarray}
L_A \  : \quad \C^n     &\rightarrow& \C^n   \nonumber \\
L_A z &=& A z , \quad z \in   \C^n ,    \label{4.2.13}
\end{eqnarray}
where $z = x + i p \ , \; (x,p) \in \R^{2n}$ and $A$ is
a complex orthogonal $n \times n$-matrix with unit determinant.
It is clear that the complex kinematic gauge theory (\ref{4.2.12})
is a complexification of the theory (\ref{4.2.1}) (see {\it Def.1}).
Besides, the standard complex structure on $\C^n$ satisfies
{\it Cond.1} and, as we will show, the map (\ref{difm}) is a bijection.
Let us demonstrate that ${\cal C}^{sat}$ is dense in
${\cal P} = \C^n$, and that in order to give the set of orbits
\begin{equation}
{\cal P} / L_{SO(n, \C)} = \C^n / L_{SO(n, \C)}     \label{4.2.14}
\end{equation}
a differentiable structure, we must exclude the origin $\{0\}$ and the orbits
of
$L_{SO(n, \C)}$ which do not intersect ${\cal C}$, that is
\begin{equation}
{\cal P}_{ph} = {\cal C} / L_{SO(n)} = Q^\C_{ph} =
{\cal P} / L_{SO(n , \C)} , \label{4.2.15}
\end{equation}
where in this equality we have implicitly excluded
non-typical orbits from ${\cal P} / L_{SO(n, \C)}$.
To show that ${\cal C}^{sat}$ is dense in
${\cal P} = \C^n$, we consider the following sets in
$\C^n$, which are invariant under $L_{SO(n, \C)}$,
\begin{equation}
D_w = \{ z \in \C^m \backslash \{ 0\} \ : \quad z^2 = w \} . \label{4.2.16}
\end{equation}
It can easily be shown that $D_0$ does not intersect $\cal C$
and that every $D_w$ (for $w \neq 0$) contains a single orbit
of $L_{SO(m, \C)}$ and  intersects ${\cal C}$. Then
\begin{equation}
{\cal C}^{sat} = \cup_{w \in \C} D_w = \C^n \backslash D_0,
\end{equation}
which proves that ${\cal C}^{sat}$ is dense
in ${\cal P} = \C^n$ since $D_0$ has real codimension
two. The orbits $D_w$ are all of the type
\begin{equation}
D_w \equiv SO(n, \C) / SO(n-1 , \C)
\end{equation}
while the orbits $\{0\}$ and $D_0$ are not.
(It can be shown that $D_0$ is just a single orbit, unless $n=2$, when it
consists of two orbits.)
Therefore, in order
to give ${\cal P} / L_{SO(n, \C)}$ a differentiable structure,
we must exclude $\{ 0\} \cup D_0$, which proves the validity of
(\ref{4.2.15}).

\subsection{Grassmann manifolds as
physical phase spaces of $U(m)$-gauge theories}

Let us consider an example with a symplectic action
of the group $U(m)$ in ${\cal P} = \R^{2mn} = T^* \R^{mn}$,
which is not the lift of an action in the configuration
space $\R^{mn}$ \footnote{We thank R. Picken for suggesting the $m=1$ example.}
. We take ${\cal P} = \R^{2mn} = \C^{mn}$,
endowed with the standard K\"ahler structure. The points
in $\cal P$ can be considered either as complex $m \times n$-matrices,
\begin{equation}
z = (z_{ri})_{r =1;}^{m;}{}_{i = 1}^{n}    \in  {\cal P}, \label{4.3.1}
\end{equation}
 or as sets of $m$ vectors in $\C^n$ (the rows of $(z_{ri})$).
Let the group $U(m)$ act by left multiplication,
\begin{equation}
L_A z = A z  , \ A \in U(m),    \label{4.3.2a}
\end{equation}
and the group $SU(n)$ by right multiplication,
\begin{equation}
L_B z =  z B  , \ B \in SU(n).    \label{4.3.2b}
\end{equation}
Both actions leave the K\"ahler structure on ${\cal P} = \C^{mn}$
invariant.

Let us assume that the group of gauge transformations
is $U(m)$. The action of $U(m)$ is free for points $z \in {\cal P} $
such that $ rank (z) = m$. Indeed, we have
\begin{equation}
L_A z = z\; \Leftrightarrow\;
\sum_{s=1}^m(A_{rs} - \delta_{rs}) z_s = 0   , \nonumber
\end{equation}
which, if $rank (z) = m$, implies that
$$
A_{rs} = \delta_{rs}  .
$$
The action (\ref{4.3.2a}) has an equivariant momentum map,
given by
\begin{eqnarray}
\mu^{\xi_{rs}}(z) &=& c \delta_{rs} - \frac{1}{4} (z_r \bar z_s +
\bar z_r z_s)  \nonumber \\
&=& c \delta_{rs} - \frac{1}{2} (p_r p_s + x_r x_s)   \label{4.3.3a} \\
\mu^{\eta_{rs}}(z) &=& \frac{i}{4} (z_r \bar z_s -
\bar z_r z_s)  \nonumber \\
&=& \frac{1}{2} (x_r p_s - x_s p_s)   , \label{4.3.3b}
\end{eqnarray}
where $u_r v_s \equiv \sum_{i=1}^{n} u_{ri} v_{si}$, and we have set
$z_r=x_r+ip_r$. The generators
$\xi_{rs}$, $\eta_{rs}$ form a basis in $u(m) \equiv Lie(U(m))$, given by
\begin{eqnarray}
\xi_{rs} &=& \frac{i}{2} (E_{rs} + E_{sr})      \nonumber \\
\eta_{rs} &=& \frac{1}{2} (E_{rs} - E_{sr}),      \label{4.3.4}
\end{eqnarray}
where the $E_{rs}$ are the elementary $m \times m$-matrices
\begin{equation}
(E_{rs})_{r's'} = \delta_{rr'}\delta_{ss'}. \nonumber
\end{equation}
One easily checks that the constant terms appearing on the right-hand side of
(\ref{4.3.3a}) are the most general constants that on cohomological grounds can
appear in the components of the momentum map.
The first-class constraints associated with this action are
\begin{eqnarray}
\mu &=& 0  \Leftrightarrow   \nonumber   \\
\mu^{\xi_{rs}} &=& 0  , \ \mu^{\eta_{rs}} =0 , \label{4.3.5}
\end{eqnarray}
or, equivalently,
\begin{equation}
z_r \bar z_s = \delta_{rs} ,  \label{4.3.6}
\end{equation}
where we have set the constant to $c = 1/2$. We see that the elements of the
constraint surface are sets of $m$ orthonormal vectors in $\C^n$
with respect to the hermitian inner product
$$
<w, w'> = w \bar w' = \sum_{i=1}^n w_i \bar w'_i  .
$$
The action of $U(m)$ on $\cal C$ is free as explained above.
Let $z^{(0)} \in {\cal C}$ and ${\cal L}$ be the subspace of
$\C^n$ spanned by $\{z_r^{(0)} \}_{r=1}^{m}$. Geometrically,
by acting with $ U(m)$ according to
\begin{equation}
(A z^{(0)})_r = A_{rs} z_{s}^{(0)},    \label{4.3.7}
\end{equation}
we obtain all the orthonormal bases of ${\cal L} \subset \C^n$.
The points in the physical phase space ${\cal P}_{ph}$ are the
orbits $[z]$ of $U(m)$ in ${\cal C}$. There is an obvious
one-to-one map between ${\cal P}_{ph}$ and the complex Grassmann manifold
$G_{m,n}(\C)$ of $m$-dimensional subspaces of $\C^n$,
\begin{eqnarray}
{\cal P}_{ph} &\rightarrow& G_{m,n}(\C)     \nonumber \\
\ [z] &\mapsto& {\cal L} = span \{ z_s \}_{s=1}^m      \label{4.3.8}
\end{eqnarray}
That this map is a diffeomorphism follows from

\begin{prop}
\begin{itemize}

\item The constraint surface $\cal C$ is an orbit in $\cal P$ under the
right action of the group $U(n)$.

\item The isotropy group is $U(n-m)$ so that
\begin{equation}
{\cal C} \stackrel{\rm diff}{\equiv} U(n)/U(n-m).
\end{equation}

\item The physical phase space ${\cal C} / U(m)$ is therefore diffeomorphic
to
\begin{equation}
{\cal P}_{ph} \stackrel{\rm diff}{\equiv}
G_{m,n}(\C) = U(n) / [U(n-m) \times U(m)]    .
\end{equation}
\end{itemize}
\end{prop}

Let us now
turn to the complexification of the symplectic action
 of $U(m)$ in order to study the form that our general
result (that the map (\ref{difm}) is a diffeomorphism)
takes in this particular example.
The complexification of $U(m)$ is
\begin{equation}
U(m)_\C = GL(m,\C).    \label{4.3.11}
\end{equation}
It is easy to verify  that the action of ${GL(m,\C)}$ on ${\cal P} = \C^{mn}$
by left multiplication
\begin{equation}
L_{A^\C} z = A^\C z  , \ A^\C \in GL(m,\C),    \label{4.3.12}
\end{equation}
is an extension of  (\ref{4.3.2a}) in the sense of (\ref{a4c}) and
satisfies the conditions (\ref{a8}) and (\ref{a9}).
Also, $0$ is a regular value of the momentum map $\mu$ so
that  (\ref{difm}) is indeed a  diffeomorphism. ${\cal C}^{sat}$
consists of all matrices $z \in \C^{mn}$ with $ rank(z) = m$.
To prove this it is sufficient to notice that for any such
matrix $z$ there is a matrix $A^\C \in GL(m,\C)$ such that
$A^\C z \in   {\cal C}$ (e.g. by using the Graham-Schmidt orthogonalization
procedure). Hence ${\cal C}^{sat}$ is dense in ${\cal P} = \C^{mn}$,
which is
in accordance with our conjecture. Furthermore, the ``bad" points outside
${\cal C}^{sat}$ are points with symmetries (i.e. with a non-trivial
isotropy subgroup of $U(m)$).

The action of $GL(m,\C)$ on ${\cal C}^{sat}$ is free by construction,
and it is clear
why the orbit space is diffeomorphic to the complex Grassmann
manifold $G_{m,n}(\C)$: the orbits are just the sets of all bases
of a given $m$-dimensional subspace of $\C^n$,
\begin{equation}
{\cal P}_{ph}=G_{m,n}(\C)={\cal C}/U(m)={\cal C}^{sat}/GL(m,\C).
\end{equation}

\vskip1.5cm

\section {Application to Yang-Mills theory}

\subsection {The hamiltonian formulation}

In this section we review some geometric properties of the infinite-dimensional
phase space of the Yang-Mills theory, which allow us to construct its
complexification along the lines proposed Sec.3 above.
We emphasize the geometric viewpoint of the theory, and refer
the interested reader to references \cite{Arm,AMM,Mon2} for the
analytic details.

Let $\Sigma$ be a compact, oriented manifold of dimension three,
$P=P(\Sigma,K)$ a principal fibre bundle over $\Sigma$, with structure
group $K$, a compact semisimple Lie group, and associated Lie algebra
$\cal K$. The Killing form in $\cal K$ will be used to
identify $\cal K$ with its dual ${\cal K}^{*}$.

The big configuration space of Yang-Mills theory is the affine space
$\bf {\cal A}$ of $\cal K$-valued connections on
$P$. For simplicity, we assume $P$ to be trivial, so that we can
identify $\bf {\cal A}$ with the affine space of
$\cal K$-valued one-forms (gauge potentials) on
$\Sigma$. (Here and in the
following all function spaces are assumed to belong to the appropriate Sobolev
classes, see \cite{Mon2} for a discussion.)
Then ${\cal A}$ is an affine space modelled on
the vector space
${\overline {\wedge}}^{1}(\Sigma ;{\cal K})$ of $\cal K$-valued one-forms
of adjoint type on $\Sigma$, and its tangent bundle can be identified with
\begin {equation}
T{\bf {\cal A}} \cong {\bf {\cal A}} \times {\overline
{\wedge}}^{1}(\Sigma ;{\cal K}).
\end {equation}
The corresponding Yang-Mills phase space will be identified with the ($L^2$-)
cotangent bundle
\begin {equation}
T^{*}{\bf {\cal A}} \cong {\bf {\cal A}} \times  {\cal X}_d (\Sigma ;{\cal K}),
\end {equation}
where ${\cal X}_d (\Sigma ;{\cal K})$ denotes the space of ${\cal
K}$-valued vector densities (or non-abelian electric fields) on $\Sigma$.
At any point $A \in {\bf {\cal A}}$, the dual
pairing between $T^{*}_A{\bf {\cal A}} \cong {\cal X}_d (\Sigma ;{\cal
K})$ and $T_A{\bf {\cal A}} \cong  {\overline
{\wedge}}^{1}(\Sigma ;{\cal K})$ is given by
\begin {equation}
(\alpha,{\tilde E}) = \int_{\Sigma} \alpha : {\tilde E},
\label{dlty}
\end {equation}
where $\alpha \in T_A{\bf {\cal A}}$, ${\tilde E} \in T^{*}_A{\bf
{\cal A}}$, and ``:" denotes the complete contraction of internal and spatial
indices (internal indices are contracted with the Killing form on $\cal K$).

The gauge group $G$ of Yang-Mills theory is the group of $K$-valued
functions $g:{\Sigma} \rightarrow K$, on $\Sigma$, and its Lie algebra
$\cal G$ is given by the Lie algebra of $\cal K$-valued functions
${\xi}:{\Sigma} \rightarrow {\cal K}$, on $\Sigma$. The dual
of ${\cal G} \cong {\wedge}^{0}(\Sigma ;{\cal K})$ is the
space ${\cal G}^{*} \cong {\wedge}_d^{0}(\Sigma ;{\cal K})$
of $\cal K$-valued scalar densities on $\Sigma$.
The dual pairing between an algebra element $\xi \in {\cal G}$ and
a scalar density ${\tilde \eta} \in {\cal G}^{*}$ is given by
\begin {equation}
\prec \xi,{\tilde \eta} \succ = \int_{\Sigma} \xi : {\tilde \eta}.
\label{dlty2}
\end {equation}
The group $G$ acts on the configuration space $\cal A$
according to the affine map
\begin {equation}
(g,A) \mapsto \varphi_g (A) = g^{-1}Ag + g^{-1}dg, \label{afma}
\end {equation}
whose canonical lift to the phase space $T^{*}\cal A$ yields the
well-known Yang-Mills transformation law
\begin {equation}
\big(g,(A,{\tilde E}) \big) \mapsto \widetilde \varphi_g (A,{\tilde E})
= \big( g^{-1}Ag + g^{-1}dg,g^{-1}{\tilde E}g \big).
\label{clbp}
\end {equation}
The action (\ref{clbp}) is symplectic with respect to
the canonical (constant) symplectic form $\Omega$ on $T^{*}\cal A$,
\begin {equation}
{\Omega}_{(A,{\tilde E})} \Big( (\delta A_1, \delta {\tilde E}_1),(\delta
A_2, \delta {\tilde E}_2) \Big) = (\delta A_1, \delta {\tilde E}_2)
- (\delta A_2, \delta {\tilde E}_1),
\end {equation}
for tangent vectors $(\delta A_i, \delta {\tilde E}_i)\in
T_{(A,{\tilde E})}(T^{*}\cal A)$, where $(\cdot,\cdot)$ denotes the
duality (\ref{dlty}).
Note that in this last equation we have used the identification
\begin {equation}
T_{(A,{\tilde E})}(T^{*}{\cal A}) \cong  {\overline
{\wedge}}^{1}(\Sigma ;{\cal K}) \times {\cal X}_d (\Sigma ;{\cal K}) .
\label{idtf}
\end {equation}

The infinitesimal generator of the configuration space action
(\ref{afma}) associated to the algebra element $\xi \in
{\cal G}$ is the vector field ${\xi}^{\cal A} \in {\cal
X{\cal A}}$, given by
\begin {equation}
{\xi}^{\cal A}(A) = (A,D_A {\xi}) = (D_A {\xi}) \frac {\delta}{\delta A},
\end {equation}
where $D_A \cdot = d \cdot + [A \wedge \cdot ]$ is the covariant
derivative defined by the connection $A$.

Analogous to the finite-dimensional case (cf. Sec.2), the phase space action
(\ref{clbp}) admits an equivariant momentum map $\mu:
T^{*}{\cal A} \rightarrow {\cal G}^{*} \cong {\wedge}_d^{0}(\Sigma
;{\cal K})$, defined as
\begin{eqnarray}
\mu (A,{\tilde E}) \bullet {\xi} &=& \mu^{\xi}(A,{\tilde E})
            = ({\xi}^{\cal A}(A),{\tilde E}) \nonumber \\
            &=& (D_A {\xi},{\tilde E})
            = \prec \xi , {\delta}_A {\tilde E} \succ,
\end{eqnarray}
where $\delta_A$ denotes the (formal) adjoint of $D_A$, and $(\cdot,\cdot)$,
$\prec\cdot
,\cdot \succ$ are the dualities (\ref{dlty}) and (\ref{dlty2}) respectively.
It follows that
$\mu (A,{\tilde E}) = {\delta}_A {\tilde E} \in {\cal G}^{*}$, and
hence the {\it Gauss constraint set} ${\cal C}={\mu}^{-1}(0)$ is
given by all phase space points satisfying the Gauss law constraint
\begin {equation}
{\delta}_A {\tilde E}=0.\label{glaw}
\end {equation}

The infinitesimal generator of the phase space action (\ref{clbp}) associated
to the algebra element $\xi \in
{\cal G}$ is the vector field ${\xi}^{T^{*} {\cal A}} \in {\cal
X({T^{*} {\cal A}}})$, given by
\begin {equation}
{\xi}^{T^{*} {\cal A}}(A,{\tilde E}) = \big((A,{\tilde E}),(D_A
{\xi},[{\tilde E},\xi])\big) = (D_A {\xi}) \frac {\delta}{\delta A} + [{\tilde
E},\xi] \frac {\delta}{\delta {\tilde E}}.
\end {equation}

\medskip
We now fix a Riemannian metric $h$ on the manifold $\Sigma$ and define an
$h$-dependent, almost-complex structure ${\bf J}$ on $T^{*}{\cal A}$
by its action on tangent vectors
\begin {equation}
{\bf J}_{(A,{\tilde E})}: (\delta A,{\delta {\tilde E}}) \mapsto (-{\delta
E}^{\flat},{\widetilde {\delta A}}^{\sharp}),
\end {equation}
using the identification (\ref{idtf}), the notation
${\delta E}^{\flat}$ for the $\cal K$-valued 1-form $h$-equi\-va\-lent to
${\delta {\tilde E}}$, and the notation ${\widetilde {\delta A}}^{\sharp}$ for
the $\cal K$-valued vector density $h$-equivalent to $\delta A$.
In complete analogy with the finite-dimensional case of equation (\ref{a9}),
this allows us to define a (weak) Riemannian metric $\gamma$ on
the cotangent bundle $T^{*}{\cal A}$ via
\begin{eqnarray}
\gamma\Big( (\delta A_1,{\delta {\tilde E}}_1),
(\delta A_2,{\delta {\tilde E}}_2)
\Big)_{(A,{\tilde E})} &:=& {\Omega}_{(A,{\tilde E})} \Big( (\delta A_1,
\delta {\tilde E}_1),{\bf J}(\delta A_2, \delta {\tilde E}_2) \Big)
\nonumber \\
&=& {\Omega}_{(A,{\tilde E})} \Big( (\delta A_1,
\delta {\tilde E}_1),(-{\delta E_2}^{\flat},{\widetilde {\delta
A}}_2^{\sharp}) \Big) \nonumber \\
&=& (\delta A_1,{\widetilde {\delta A}}_2^{\sharp}) +
({\delta E_2}^{\flat},\delta {\tilde E}_1).\label{wrie}
\end{eqnarray}
Note that the almost-complex structure $\bf J$, and therefore also
the Riemannian metric $\gamma$ are $G$-invariant.
Since the actions (\ref{afma}) and (\ref{clbp}) are affine,
their differentials are the respective linear parts, and hence
\begin{eqnarray}
{\bf J} \circ d{
\widetilde \varphi}_g(\delta A,{\delta {\tilde E}}) &=& {\bf J}
(g^{-1}\delta A g,g^{-1}{\delta {\tilde E}}g) \nonumber \\
&=& ( g^{-1}{\delta E}^{\flat}g,g^{-1}{\widetilde {\delta A}}^{\sharp}g )
\nonumber \\
&=&  d{\varphi}_g \circ {\bf J}(\delta A,{\delta {\tilde E}}),
\end{eqnarray}
proving the $G$-invariance. The situation is formally the same
as in Sec.3 above, i.e. we have a $G$-invariant almost-K\"ahler
structure $({\bf J},\gamma)$ obeying the fundamental relation
\begin {equation}
\gamma (\cdot,\cdot) = \Omega (\cdot,{\bf J}\cdot).
\end {equation}

Considering as before the tangent map
$d{\mu}$ and its ($L^2$-)adjoint $\nabla \mu = d{\mu}^{*}$, using
elliptic theory \cite{Arm,AMM} we once more deduce the (gauge-invariant)
Moncrief decomposition (\ref{moncr}), this time of the tangent space
$T_p(T^{*}{\cal A})$, at any
point $p=(A_o,{\tilde E}_o)$ of the Gauss constraint set ${\cal C}$.

In order that ${\cal C}$ be a manifold, we must exclude from it
points $(A,\tilde E)$ which possess one or more so-called infinitesimal
symmetries, i.e. covariantly constant, non-zero functions $\xi \in {\cal
G} \cong {\wedge}^{0}(\Sigma ;{\cal K})$ that commute with the
electric field $\tilde E$,
\begin {equation}
D_A \xi = 0 \ \ \ \ {\rm and} \ \ \ \ [{\tilde E},\xi] = 0.
\end {equation}
In a point $(A,\tilde E)$ with infinitesimal gauge symmetries, the
Yang-Mills field variables can be reduced to take their values in the Lie
algebra of a smaller group $H\subset K$.
Note that the kernel $Ker \,  {\nabla \mu}_{(A_o,{\tilde
E}_o)}$ coincides with the set of infinitesimal symmetries of the configuration
${(A_o,{\tilde E}_o)}$.
One can show that $Ker \, {\nabla \mu}_{(A_o,{\tilde
E}_o)}=\{0\}$ iff ${d \mu}_{(A_o,{\tilde E}_o)}$ is surjective which, by the
implicit function theorem, implies the following result \cite{AMM}:

\begin{theo}
If $(A_o,{\tilde E}_o)$ has no infinitesimal symmetries, then the
Gauss constraint set ${\cal C}={\mu}^{-1}(0)$
is a manifold near $(A_o,{\tilde
E}_o)$, with tangent space given by $Ker \, d{\mu}_{(A_o,{\tilde E}_o)}$.
\end{theo}

We therefore exclude all points with symmetries from the Gauss constraint
surface, and continue to denote the resulting
smooth manifold by ${\cal C}$. In
every point $(A_o,{\tilde E}_o)$ of this manifold, the operator ${\nabla
\mu}_{(A_o,{\tilde E}_o)}$ has a trivial kernel and the ``Laplacian" operator
$\Delta = {d \mu}_{(A_o,{\tilde
E}_o)} \circ {\nabla \mu}_{(A_o,{\tilde
E}_o)}$ is an isomorphism from  ${\cal G} \cong {\wedge}^{0}(\Sigma
;{\cal K})$ to ${\cal G}^{*} \cong {\wedge}_d^{0}(\Sigma
;{\cal K})$. Marsden-Weinstein reduction leads to
the reduced, physical phase space ${\cal P}_{ph} = {\cal C}/G$,
which is symplectomorphic to the cotangent bundle
$T^{*}{\cal M}_{ph}$ of the moduli space ${\cal M}_{ph}=
{\cal A}'/G$ of equivalence classes of non-symmetric
connections ${\cal A}'$. Hence all aspects of the finite-dimensional
kinematic discussion of Sec.2 are realized in the field-theoretic example
of the Yang-Mills phase space.

\subsection {Complexification}

Let us denote by ${\cal A}^{\C}$ the complex affine space of
${\cal K}_\C$-valued
connection one-forms on $\Sigma$ (${\cal K}_\C$ is the Lie algebra of
the universal complexification $K_\C$
of $K$ introduced in Sec.3). ${\cal A}^{\C}$ is the complexification of
the real affine space $\cal A$, and is modelled on the vector space
${\overline {\wedge}}^{1}(\Sigma ;{\cal K}_{\C})$.
The choice of a Riemannian metric $h$ on
$\Sigma$ provides us with a one-to-one
map ${\cal A}^{\C} \rightarrow T^{*}{\cal A} $, defined by

\begin {equation}
A+i{E} \mapsto (A,{\tilde E}^{\sharp})\label{cmap}
\end {equation}
where ${\tilde E}^{\sharp}$   denotes the $\cal K$-valued vector density
$h$-equivalent to the $\cal K$-valued one-form $E$
\footnote{Since $A$ and $E$ have different physical dimension,
we should be writing $A+iLE$ instead of $A+iE$ in (\ref{cmap}), where $L$ is
a fixed constant with the dimension of length. For simplicity we have set
this constant equal to one.}. This is well defined,
since $A+iE$ is a $K_\C$-valued one-form of $Ad_{K_\C}$-type, and in
particular of $Ad_{K}$-type. This implies that (restricting to the subgroup
$K\subset K_\C$) $A$ is
$K$-pseudotensorial and ${\tilde E}^{\sharp}$ of $Ad_{K}$-type.

Following the construction given in Sec.3, we now ``complexify" the
infinitesimal ${\cal G}$-action on $T^{*}{\cal A}$,
\begin {equation}
\xi \mapsto
{\xi}^{T^{*} {\cal A}}(A,{\tilde E}) = (D_A {\xi}) \frac {\delta}{\delta A}
+ [{\tilde E},\xi] \frac {\delta}{\delta {\tilde E}},\label{gact1}
\end {equation}
by representing the imaginary generators according to
\begin {equation}
i{\xi} \mapsto {\bf J}{\xi}^{T^{*}{\cal A}}.\label{gact2}
\end {equation}

Let us verify that the infinitesimal $G_{\C}$-gauge action on
the complex Yang-Mills {\it configuration} space ${\cal A}^{\C}$ is
compatible, under the map (\ref{cmap}), with the infinitesimal ``complexified"
action on $T^{*}{\cal A}$, given by (\ref{gact1}) and (\ref{gact2}) (i.e. that
{\it Cond.1} of Sec.3 holds). For the former, one has
\begin{eqnarray}
(\xi +i \eta) \cdot (A+iE) &=& D_{A+iE}(\xi +i\eta) \nonumber \\
&=& d(\xi +i\eta) + [(A+iE),(\xi +i\eta)] \nonumber \\
&=& D_A\xi - [E,\eta] +i(D_A \eta + [E,\xi]),\label{gact4}
\end{eqnarray}
while for the latter one derives
\begin{eqnarray}
(\xi +i \eta) \cdot (A,{\tilde E}^{\sharp}) &=& \xi \cdot (A,{\tilde
 E}^{\sharp})
+ {\bf J} \big( \eta \cdot (A,{\tilde E}^{\sharp}) \big) \nonumber \\
&=&\big( D_A \xi, [{\tilde E}^{\sharp},\xi] \big) + {\bf J} \big( D_A \eta,
[{\tilde E}^{\sharp},\eta] \big) \nonumber \\
&=& \big( D_A \xi - [E,\eta], [{\tilde E}^{\sharp},\xi] + {\widetilde {D_A
\eta}}^{\sharp} \big).\label{gact5}
\end{eqnarray}
Clearly, the right-hand side of (\ref{gact4}) is mapped into the
right-hand side of (\ref{gact5}) under the map (\ref{cmap}). We have
therefore verified the claim made in Sec.3 above, that the kinematic,
complex Yang-Mills theory with structure group $K_\C$ satisfies
{\it Def.1} and {\it Cond.1} with respect to its real counterpart with
structure group $K$.

The validity of the Moncrief decomposition (\ref{moncr}) and the inverse
mapping theorem (in the appropriate
weighted Sobolev spaces \cite{Mon2}), and the fact that (\ref{wrie}) is a weak
Riemannian metric in $T^*{\cal A}$ imply that {\it Prop.1} and {\it Prop.2} of
Sec.3 are valid in the present case and therefore the
following theorem (the analogue of {\it Th.3} for the infinite-dimensional
case of a Yang-Mills theory) holds:

\begin {theo}
Consider a Yang-Mills theory with compact structure group
$K$ corresponding to the trivial bundle
$P(\Sigma,K)$ over the three-dimensional oriented and compact
manifold $\Sigma$. Then the map (\ref{difm}) is a diffeomorphism
between the physical phase space ${\cal C}/ \widetilde L_G$
of the Yang-Mills theory with structure group $K$
and an open submanifold ${\cal C}^{sat} / L_{G_\C}$
of the physical configuration space of the Yang-Mills theory
with structure group $K_\C$ (assuming that points with symmetries
have been excluded from $\cal C$).

\end {theo}

\subsection{Holomorphic Wilson loops}

With these ingredients in hand, we can now form the saturation ${\cal C}
^{sat}$ of the Gauss constraint manifold ${\cal C}$
according to formula (\ref{csat}). If we can show that any
$G_\C$-orbit
through a point $A^\C\in {\cal A}^\C$ cuts the constraint surface
${\cal C}$, we have ${\cal C}^{sat}\cong {\cal A}^\C$, and
thus an equivalence of the quotient spaces
\begin{equation}
{\cal C}/G \cong {\cal A}^\C/G_\C.
\end{equation}

Note that, due to the geometric structures available on ${\cal A}^\C$,
we do never have the problem of uniqueness of the ``gauge choice"
${\delta}_A {\tilde E}=0$, the Gauss law constraint of equation
(\ref{glaw}), i.e. a given $G_\C$-orbit can never cut the surface
${\cal C}$ more than once. For the attainability of this ``gauge
choice", one derives the following conditions. Starting from an
arbitrary point $A^\C=A +iE\in {\cal A}^\C$, one looks for a complex
gauge transformation $g_\C=g\cdot\exp i\omega\in G_\C$, such that
the configuration $(A',\tilde E')$, where
$\varphi_{g_\C}(A+iE)=A' +iE'$, lies in
${\cal C}$. Taking w.l.o.g. $g_\C$ to be purely imaginary,
$g_\C=e^{i\omega}$, one derives the following non-linear equations for
$\omega$:
\begin{eqnarray}
A'&=&\sum_{n=0}^\infty \frac{(-1)^n}{(2n)!} (Ad\,\omega)^{2n} A-
 \sum_{n=0}^\infty \frac{(-1)^{n+1}}{(2n+1)!} (Ad\,\omega)^{2n+1}(E+d\omega),
\nonumber \\
E'&=&\sum_{n=0}^\infty \frac{(-1)^{n+1}}{(2n+1)!} (Ad\,\omega)^{2n+1} A+
 \sum_{n=0}^\infty \frac{(-1)^n}{(2n)!} (Ad\,\omega)^{2n} (E+d\omega),
\label{nonl}
\end{eqnarray}
using the notation
\begin{equation}
(Ad\,\omega)^n X := [\omega,[\omega,\dots [\omega,[\omega,X]]\dots ]],\quad
(Ad\,\omega)^0 X := X.
\end{equation}

Comparing with the finite-dimensional $SO(n)$-gauge model, one may not
expect the equations (\ref{nonl}) to possess solutions for arbitrary
$(A,E)$, but still these relations may be useful in determining whether
${\cal C}^{sat}$ is dense in ${\cal A}^\C$, as we are
conjecturing.

Recall there is a natural set of gauge-invariant variables on any space
$\cal A$ of connections, given by the so-called {\it Wilson loops}
\begin{equation}
T_\gamma (A):= Tr\,{\rm P}\exp\oint_\gamma A,
\label{wils}
\end{equation}
where $\gamma$ is a closed curve in $\Sigma$, and P denotes path-ordering
along $\gamma$. The expression ${\rm P}\exp\oint_\gamma A$ is also known
as the holonomy of the connection $A$ along the loop $\gamma$.

It is well known that for compact structure group $K$, the knowledge of the
values of all Wilson loops $T_\gamma$ is equivalent to the knowledge of the
gauge connection $A$ up to gauge transformations \cite{Gil}, and the
variables $\{T_\gamma\}$ form an overcomplete set of coordinates on the
physical configuration space ${\cal A}/G$. For non-compact $K$, the
Wilson loops may not be completely separating, i.e. there may be sets of
connection configurations from ${\cal A}/G$ that are mapped into the
same $T_\gamma$-configuration.
However, for the case $K_\C=SL(2,\C)=SU(2)_\C$, which is the one
relevant for the Ashtekar formulation of $3+1$ gravity, the points in
${\cal A}^\C/G_\C$ which are not separated by the Wilson loop variables form a
set of measure zero \cite{Gold}, and moreover the Wilson loops
separate all points which can be separated in the non-Hausdorff space
${\cal A}^\C/G_\C$ \cite{AshLew}.

Let us now consider the traced holonomies (\ref{wils}) as functions on the
space ${\cal A}^\C/ G_\C$ of complex gauge connections modulo gauge
transformations introduced earlier (see \cite{RovSmo}).
As mentioned above these variables have been proven in
\cite{Gold,AshLew} to form a set of good generalized coordinates
on the physical {\it configuration} space of the Yang-Mills theory
with complex structure group $K_\C = SL(2, \C)$. Our {\it
Props.1} and {\it 2} and {\it Th.5}
provide a set of necessary conditions for the variables (\ref{wils})
to be good generalized coordinates on the physical {\it phase}
space of the Yang-Mills theory with structure group $SU(2)$.
(A proof that ${\cal C}^{sat}$ is dense would provide a sufficient condition.)
This is also relevant to general relativity written in terms of
Ashtekar variables, since there ${\cal A}^\C$
 is the (big) phase space rather
than the (big) configuration space of the theory. Although the
symplectic structure of general relativity is very different from
that of SU(2) Yang-Mills
theory, one may expect that an analogue of our result is valid also in the
case of gravity. This would indicate that (\ref{wils}) are
good generalized coordinates on the phase space of general relativity in
the Ashtekar formulation (complementing the results of \cite{Gold,AshLew}).
A proof of this assertion would however involve a proof
of {\it Th.5} and of the conjecture that ${\cal C}^{sat}$
is dense also for the gravity case, which is beyond
the scope of the present paper.

On the other hand, our result points to an alternative
to the usual loop space formulations of
Yang-Mills theory, where the Wilson loops (\ref{wils}) are used as
variables on the physical {\it configuration} space, and where generalized
Wilson loops have to be introduced in a rather asymmetric way to bring in the
dependence on the canonically conjugate momenta $\tilde E$ \cite{GamTri}.

A next important step in our construction is the search for natural
algebraic structures on the set of holomorphic and anti-holomorphic
Wilson loops, which could
serve as a starting point for the quantization.
\vskip1.5cm

\section{Conclusions}

We have derived a number of conditions, under which it is possible to obtain
an alternative description for the reduced phase space of a hamiltonian
first-class constrained system. Instead of using the two-step
Marsden-Weinstein reduction associated with the group $G$ of gauge
transformations, one takes a single quotient with respect to an appropriate
phase space action of the complexified group $G_\C$
({\it Th.3} and {\it Th.5}). A necessary condition
for these two methods to lead to equivalent results is the existence of an
appropriate extension of the $G$-action in the directions perpendicular to
the constraint surfaces.

We conjecture that the conditions for equivalence we establish are
actually sufficient, and hence that the saturation ${\cal C}^{sat}$ (all points
in phase space that can be reached from the zero-momentum constraint surface
$\cal C$ by a complex gauge transformation from $G_\C$) is dense in the phase
space. This conjecture is corroborated by all the finite-dimensional examples
investigated in Sec.4, but a general proof is still lacking. The examples
with $G=SO(n)$ and $G=U(m)$ also demonstrate that no obstructions in principle
occur when the $G$-action is non-free or the phase space not of the form
of a cotangent bundle over the configuration space.

Due to the dual interpretation of the space underlying our construction
(either as the phase space $\cal P$
of a (real) gauge system or as the configuration space $Q^\C$ of
a (complex) gauge system), it allows for a variety of physical applications,
some of which were mentioned in the introduction. In any case one has to
show that a (hamiltonian or lagrangian) dynamics can be introduced into the
framework in a consistent way. We leave the discussion of this and further
applications to a future publication.

\medskip

{\it Acknowledgements.} We are grateful to A. Ashtekar, R. Picken and
A.R. Silva for useful discussions and to J. Rawnsley for a comment. JNT and JMM
were partially supported by JNICT grant PESO/P/PRO/8/91 and GTAE, and RL
partially by a DFG scholarship.

\end {document}